\documentclass[preprintnumbers,amssymb,amsmath,nofootinbib,showpacs]{revtex4}
\usepackage{graphicx}
\usepackage{dcolumn}
\usepackage{bm}
\usepackage{comment}
\usepackage{color}
\usepackage[utf8]{inputenc}
\usepackage{amsmath}
\usepackage{hyperref}
\usepackage{bbm}
\begin{document}

\title{Effect of Transition Magnetic Moments on Collective Supernova Neutrino Oscillations}

\author{Andr\'e de Gouv\^ea}
\affiliation{Department of Physics \& Astronomy, Northwestern University, IL 60208-3112, USA}
\author{Shashank Shalgar}
\affiliation{Department of Physics \& Astronomy, Northwestern University, IL 60208-3112, USA}

\pacs{14.60.Pq, 97.60.Bw}

\preprint{NUHEP-TH/12-06}
\date{\today}

\begin{abstract}
We study the effect of Majorana transition magnetic moments on the flavor evolution of neutrinos and antineutrinos inside the core of Type-II supernova explosions. We find non-trivial collective oscillation effects relating neutrinos and antineutrinos of different flavors, even if one restricts the discussion to Majorana transition electromagnetic moment values that are not much larger than those expected from standard model interactions and nonzero neutrino Majorana masses. This appears to be, to the best of our knowledge, the only potentially observable phenomenon sensitive to such small values of Majorana transition magnetic moments. We briefly comment on the effect of Dirac transition magnetic moments and on the consequences of our results for future observations of the flux of neutrinos of different flavors from a nearby supernova explosion.
\end{abstract}

\maketitle

\section{Introduction}

The understanding of neutrino flavor oscillations in the presence of a dense neutrino background has improved dramatically over the last several years \cite{Pantaleone:1992eq,Samuel:1993uw,Qian:1995ua,Pastor:2002we,Sawyer:2005jk,Duan:2005cp,Hannestad:2006nj,Duan:2007mv,Raffelt:2007yz,EstebanPretel:2007ec,Raffelt:2007cb,Raffelt:2008hr}. 
The collective effects mediated by neutrino self-interactions lead to new oscillation phenomena, some of which may be imprinted, in an observable way, in the flavor and energy spectra of neutrinos from Type-II supernova explosions. 

\begin{figure}
\includegraphics[width=8.5cm]{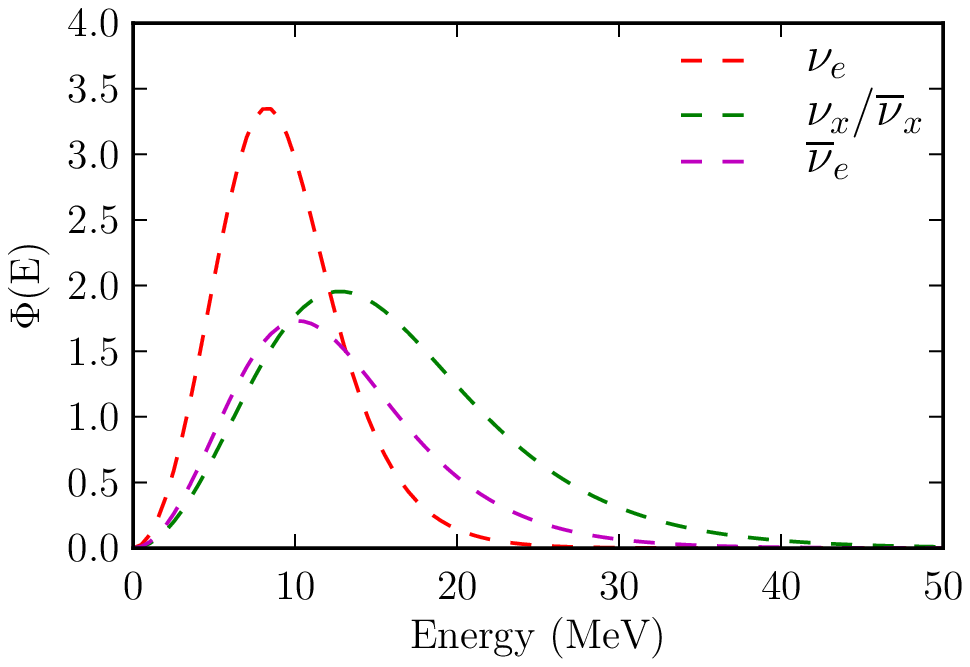}
\includegraphics[width=8.5cm]{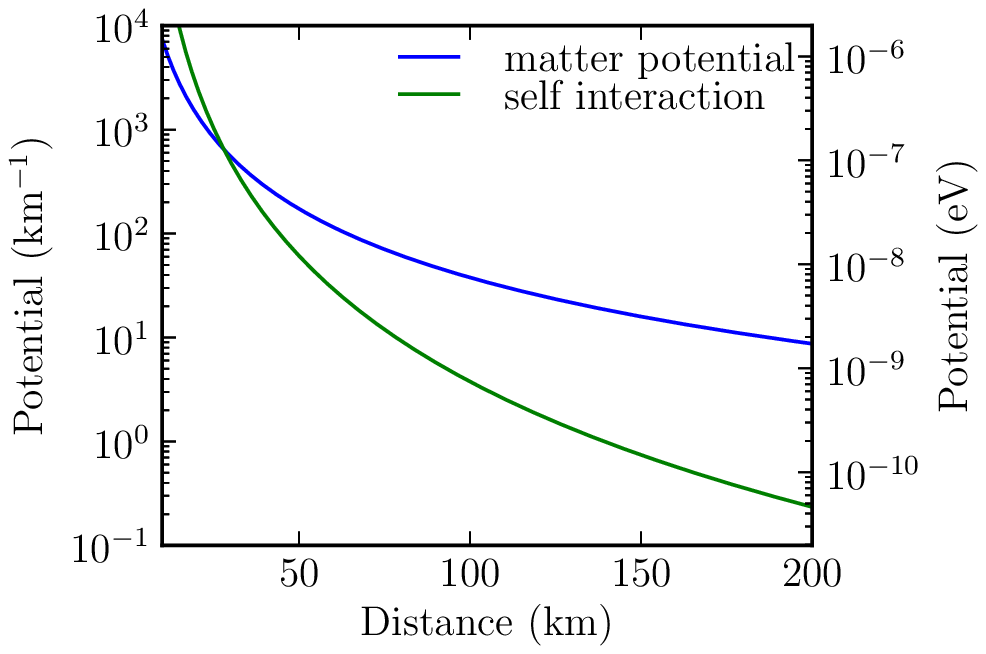}
\caption{(Left) Initial $\nu_e, \bar{\nu}_e, \nu_x, \bar{\nu}_x$ neutrino fluxes as a function of the neutrino energy, and (right) the strength of the matter potential and the effective neutrino self-interaction as a function of the distance from the center of the supernova explosion. The matter and self-interaction potentials are defined as $\sqrt{2}G_{F}n_{e}$ and $\sqrt{2}G_{F}n_{\nu}$, respectively. See Section~\ref{sec:formalism} for details.}
\label{pot:ini}
\end{figure}

\begin{figure}
\includegraphics[width=8.5cm]{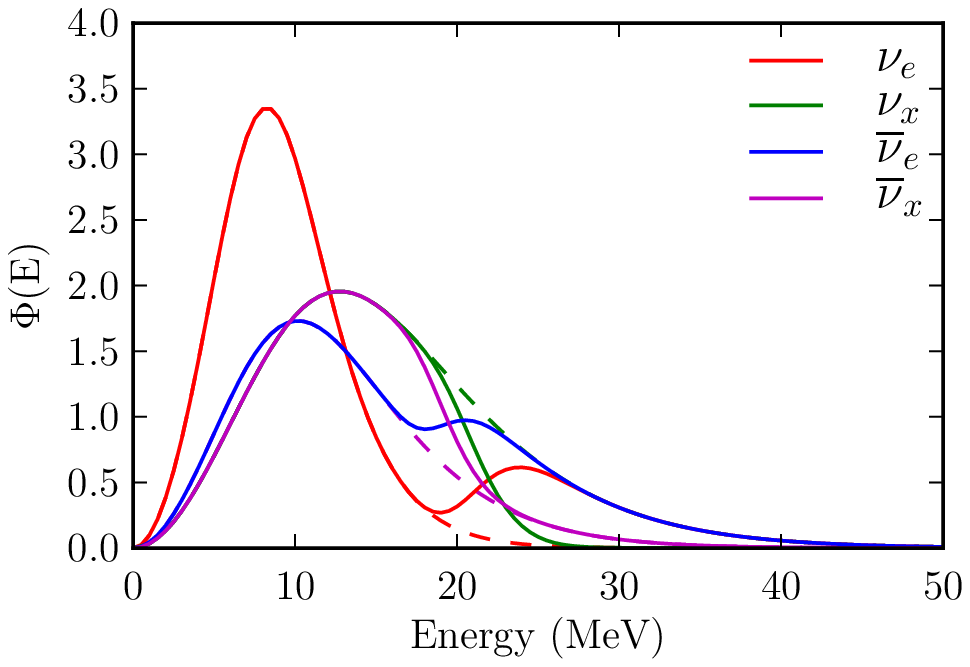}
\includegraphics[width=8.5cm]{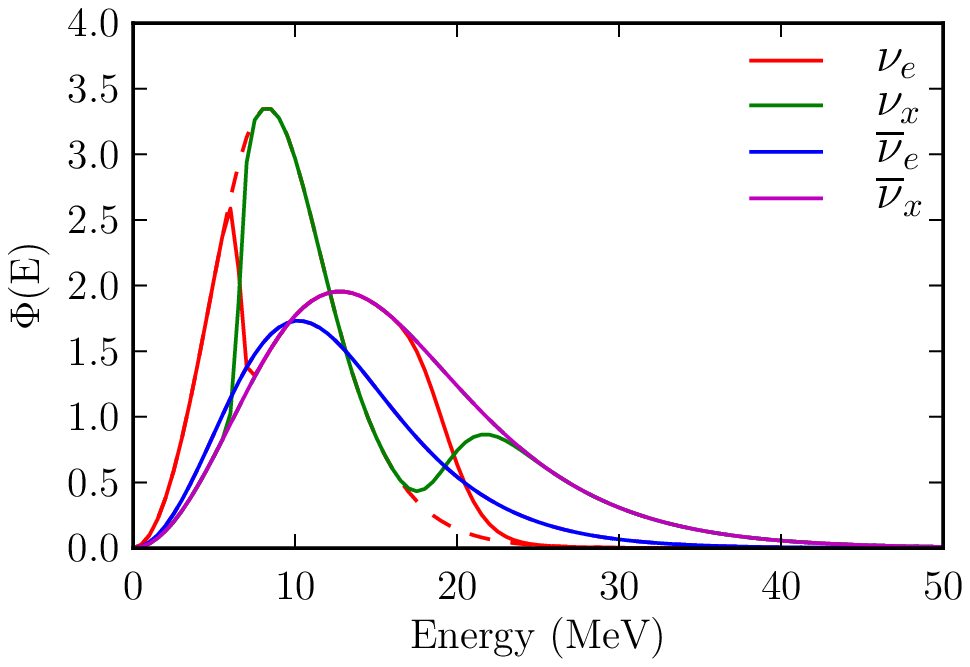}
\caption{Initial (dashed, $r=50$~km) and final (solid, $r=200$~km) $\nu_e, \bar{\nu}_e, \nu_x, \bar{\nu}_x$ neutrino fluxes as a function of the neutrino energy for the normal (left) and inverted mass hierarchy (right), for $|\Delta m^2|=2.5\times 10^{-3}$~eV$^2$ and $\sin^2\theta=10^{-4}$. See~\cite{youtube:list} for an animation of the radial evolution of the fluxes.}
\label{stable:1}
\end{figure}

These novel effects are very sensitive to some of the currently unknown neutrino masses and mixing parameters, in particular the neutrino mass hierarchy. For example, in the case of the so-called inverted neutrino mass hierarchy \cite{Dasgupta:2008my,Chakraborty:2008zp,Gava:2009pj},  it has been established that there is a ``flavor spectral split'' in the neutrino flux \cite{Fogli:2007bk,Duan:2006an,Raffelt:2007xt,Duan:2007fw,Duan:2007bt,Fogli:2008pt,Dasgupta:2009mg,Duan:2010bg}, while none is observed in the case of the normal neutrino mass hierarchy. Using initial $\nu_e, \bar{\nu}_e, \nu_x, \bar{\nu}_x$ neutrino fluxes as a function of the neutrino energy from the steep power law model computed in \cite{Keil:2002in} (with $p=10$ and $q=3.5$) and depicted in  Fig.~\ref{pot:ini}(left), and assuming the matter potentials are as depicted in Fig.~\ref{pot:ini}(right), we reproduce these results under a two-flavor, single-angle approximation in Fig.~\ref{stable:1}. 
See \cite{Dasgupta:2008cd,EstebanPretel:2007yq,Dasgupta:2007ws,Duan:2008za,Fogli:2008fj,Friedland:2010sc,Dasgupta:2010cd,Duan:2010bf} and \cite{Sawyer:2008zs,Duan:2008eb,Duan:2008fd,Duan:2010bg} for three-flavor and multi-angle analyses of the problem.
Our results, which will be discussed in more detail in Section~\ref{sec:formalism}, are in good agreement with previous calculations that make use of the same initial fluxes and matter profile \cite{Duan:2010bg,Chakraborty:2008zp}. 
While the split in the neutrino spectrum in the case of an inverted hierarchy is present only for a non-vanishing mixing angle $\theta$, its location and shape are not very sensitive to the actual value of $\theta$. This is a testament to the fact that due to the high degree of non-linearity there are ``switch-on'' effects for certain parameters in the Hamiltonian. 

The recent discovery that the neutrino mixing angle $\theta_{13}$ \cite{Nakamura:2010zzi} is not zero \cite{Abe:2011sj,Adamson:2011qu,Abe:2011fz,An:2012eh,Ahn:2012nd} implies that something similar to the phenomenon depicted in Fig.~\ref{stable:1} is expected from supernova neutrinos. It should be noted, however, that while the energy spectra of neutrinos are very different for the two hierarchies, it is not yet entirely clear whether a robust neutrino mass hierarchy measurement with supernova neutrinos is practical considering the uncertainties in the initial fluxes \cite{Choubey:2010up}. Here  we will not concern ourselves with this very important practical issue.

In this paper we study the effect of nonzero neutrino magnetic moments in the presence of a strong magnetic field on collective neutrino oscillations. The effect of neutrino magnetic moments in neutrino oscillations in variable matter density is an interesting phenomenon that was first identified as a potential solution to the solar neutrino puzzle several years ago \cite{Lim:1987tk,Akhmedov:1988uk}, but its consequences for collective effects in supernova neutrino oscillations are, to the best of our knowledge, yet to be discussed in the literature. 

We assume that neutrinos are Majorana fermions. When the standard model is augmented to include nonzero Majorana neutrino masses and lepton mixing, electroweak interactions lead to very small but nonzero transition magnetic moments. These results are summarized in Section~\ref{unknown:1}, along with our assumptions regarding the magnetic field inside the supernova explosion.  We limit our discussion to Majorana neutrinos for several reasons. Some are related to technical issues concerning numerical calculations:  Majorana neutrinos only have nonzero transitional magnetic moments and hence there are fewer free parameters, and only the transverse component of the magnetic field affects their flavor evolution.  More important, in the case of Majorana neutrinos, the magnetic moment interactions with the magnetic field mediate ``neutrino--antineutrino oscillations'' and one naively anticipates the possibility of spectral splits between neutrinos and antineutrinos of different flavors. Our results, presented in Section~\ref{sec:results}, confirm this suspicion and reveal that even very small values the magnetic moment, perhaps those close to the ``standard model'' level, may play a significant role in the evolution of neutrino fluxes due to the high degree of non-linearity in the problem. 

In our attempt to draw attention to the importance of the transition magnetic moment (and in order to simplify the discussion), throughout, we are limiting our discussion to the so-called single angle approximation, and pretend there are only two neutrino/antineutrino flavors: $\nu_e$ and $\nu_x$, the latter a linear combination of $\nu_{\mu}$ and $\nu_{\tau}$. We comment on the inclusion of multi-angle and three-flavor effects in the conclusions, Section~\ref{sec:conclusion}. There we also address other simplifying assumptions, and comment on our expectations regarding the impact of magnetic moments if the neutrinos are Dirac fermions.

\section{Collective Neutrino Oscillations}
\label{sec:formalism}

We discuss neutrino oscillations inside the supernova using the density matrix formulation. The evolution of the neutrino states is given by
\begin{equation}
i\dot{\rho}=[\rho,H].
\label{eq:time}
\end{equation}
We take $\rho$ to be a $4 \times 4$ matrix\footnote{In the absence of effects that mix neutrinos and antineutrinos, it is customary (see, for example, \cite{Sigl:1992fn}), and more convenient, to describe two-flavor oscillations, including antineutrinos, with a $2 \times 2$ density matrix.} which includes both ``neutrino'' (left-handed) and ``antineutrino'' (right-handed) states, and $H$ is the Hamiltonian, which governs the time evolution of the system. This will prove useful when we consider the effect of the transition magnetic moment. 

We define the density matrix $\rho$ in the flavor basis such that its initial state is diagonal:
\begin{equation}
\rho=
\begin{pmatrix}
\rho_{ee} & \rho_{ex} & \rho_{e\bar{e}} & \rho_{e\bar{x}} \\
\rho_{x e} & \rho_{xx} & \rho_{x\bar{e}} & \rho_{x\bar{x}} \\
\rho_{\bar{e} e} & \rho_{\bar{e}x} & \rho_{\bar{e}\bar{e}} & \rho_{\bar{e}\bar{x}} \\
\rho_{\bar{x} e} & \rho_{\bar{x}x} & \rho_{\bar{x}\bar{e}} & \rho_{\bar{x}\bar{x}} 
\end{pmatrix}.
\label{rho:1}
\end{equation}
$\rho_{ij}$ are defined as follows:
\begin{equation}
(2\pi)^{3}\delta^{(3)}(\mathbf{p}-\mathbf{p^{\prime}})\rho_{ij}(\mathbf{p}) =
\begin{cases}
\langle a_{j}^{\dagger}(\mathbf{p}) a_{i}(\mathbf{p^{\prime}}) \rangle & \text{if }i,j=e,x, \\[0.5ex] 
\langle b_{j}^{\dagger}(\mathbf{p}) b_{i}(\mathbf{p^{\prime}}) \rangle & \text{if }i,j=\bar{e},\bar{x}, \\[0.5ex]
\langle a_{j}^{\dagger}(\mathbf{p}) b_{i}(\mathbf{p^{\prime}}) \rangle & \text{if }i=\bar{e},\bar{x} \text{ and } j=e,x, \\[0.5ex]
\langle b_{j}^{\dagger}(\mathbf{p}) a_{i}(\mathbf{p^{\prime}}) \rangle & \text{if }i=e, x \text{ and } j=\bar{e},\bar{x}.
\end{cases}
\label{creation_op}
\end{equation}
We use the notation of \cite{Sigl:1992fn} and denote the creation and annihilation operators for neutrinos (anti-neutrinos) by $a^{\dagger}$($b^{\dagger}$) and $a$($b$) respectively.\footnote{More precisely, we also extend the notation in \cite{Sigl:1992fn} to include the ``neutrino--antineutrino'' elements (block--off-diagonal components) of the density matrix.} We also define $\rho^c$ to be equal to $\rho$ with $a$ replaced by $b$ and vice-versa,
\begin{equation}
\rho^{c}=
\begin{pmatrix}
\rho_{\bar{e}\bar{e}} & \rho_{\bar{e}\bar{x}} & \rho_{\bar{e}e} & \rho_{\bar{e}x} \\
\rho_{\bar{x} \bar{e}} & \rho_{\bar{x}\bar{x}} & \rho_{\bar{x}e} & \rho_{\bar{x}x} \\
\rho_{e\bar{e}} & \rho_{e\bar{x}} & \rho_{ee} & \rho_{ex} \\
\rho_{x \bar{e}} & \rho_{x\bar{x}} & \rho_{x e} & \rho_{xx} 
\end{pmatrix}.
\label{rhoc:1}
\end{equation}
The density matrix $\rho^{c}$ can be obtained from $\rho$ by replacing the blocks $11 \leftrightarrow 22$ and $12 \leftrightarrow 21$.

The Hamiltonian $H$ will be  divided into three components: $H=H_{vac}+H_{mat}+H_{self}$. $H_{self}$, which contains the neutrino--neutrino self-interactions, is
\begin{equation}
H_{self} = \sqrt{2}G_{F}n_{\nu} \int dE ~ G^{\dagger}(\rho(E) - \rho(E)^{c*})G+\frac{1}{2}G^{\dagger}\mathrm{Tr}\left((\rho(E)-\rho(E)^{c*})G\right),
\label{h:self}
\end{equation}
where $n_{\nu}$ is the total effective number of neutrinos (number of neutrinos plus the number of antineutrinos), $G_F$ is the Fermi constant, and $G$ is a matrix of dimensionless coupling constants,
\begin{equation}
G=
\begin{pmatrix}
1 & 0 & 0 & 0 \cr
0 & 1 & 0 & 0 \cr
0 & 0 & -1 & 0 \cr
0 & 0 & 0 & -1 
\end{pmatrix}.
\end{equation}
A derivation of Eq.~\ref{h:self}, along with some of the notation, is discussed in Appendix~\ref{app:1}. Note that we have a complex-conjugation in the second component of Eq.~(\ref{h:self}),\footnote{Using the fact that $\rho$ is Hermitian, it is easy to see that $\rho_{ee}^{*}=\rho_{ee}$, $\rho_{\bar{e}x}^{*}=\rho_{x\bar{e}}$, etc.} which is different from the convention of $\rho$ defined in \cite{Sigl:1992fn}. 

The interactions of neutrinos with the rest of the environment are contained in
\begin{equation}
H_{mat} = 
\begin{pmatrix}
\sqrt{2}G_{F}(n_{e}-\frac{n_{n}}{2}) & 0 & 0 & 0\\
0 & -\sqrt{2}G_{F}\frac{n_{n}}{2} & 0 & 0\\
0 & 0 & -\sqrt{2}G_{F}(n_{e}-\frac{n_{n}}{2}) & 0\\
0 & 0 & 0 & \sqrt{2}G_{F}\frac{n_{n}}{2}\\
\end{pmatrix}.
\end{equation}
This term is the same as the usual MSW term, written in the form of a $4 \times 4$ matrix. $n_{e,n}$ are, respectively, the electron and neutron number densities. The neutral current interactions can in principle play a role in the presence of lepton number violation as there is the possibility of neutrino--antineutrino oscillations. Henceforth, however, we neglect them as they do not qualitatively impact our results and would only render the discussion less clear.

Finally, the relation between the initial states and the vacuum Hamiltonian eigenstates is contained in the ``vacuum'' part of the Hamiltonian, given by (in the $4\times 4$ form used here), in the absence of magnetic fields,
\begin{equation}
H_{vac} = 
\begin{pmatrix}
-\omega \cos 2\theta & \omega \sin 2\theta  & 0 & 0\cr 
\omega \sin 2\theta & \omega \cos 2\theta  & 0 & 0\cr
 0 & 0 & -\omega \cos 2\theta  & \omega \sin 2\theta \cr 
 0 & 0 & \omega \sin 2\theta & \omega \cos 2\theta
\end{pmatrix},
\label{hvac:1}
\end{equation}
where $\omega=\frac{\Delta m^{2}}{4E}$ is the vacuum oscillation frequency, and $\theta$ is the mixing angle that relates the weak and flavor bases. For all practical purposes, $\theta$ is to be interpreted as the lepton mixing angle $\theta_{13}$ and $\Delta m^{2}$  as $\Delta m^2_{13}\sim\Delta m^2_{23}$ \cite{Nakamura:2010zzi}. Throughout, unless otherwise noted, we fix $\theta=10^{-2}$ radians, in spite of the recent evidence for a much larger $\theta_{13}$. This is done in order to focus on the switch-on effect due to collective oscillations and in order to suppress any additional structure due to a ``large'' mixing angle. Figure~\ref{stable:1} was generated by numerically solving Eq.~(\ref{eq:time}) for the $4\times 4$ density matrix as a function of time using the initial conditions and matter profiles depicted in Figure~\ref{pot:ini}.

In the presence of a nonzero Majorana transition magnetic moment and a background magnetic field, the ``vacuum'' Hamiltonian is modified~(see, e.g., \cite{Giunti:2008ve}): 
\begin{equation}
H_{vac} = 
\begin{pmatrix}
-\omega \cos 2\theta & \omega \sin 2\theta  & 0 & \mu B_{T}\cr 
\omega \sin 2\theta & \omega \cos 2\theta  & -\mu B_{T} & 0\cr
 0 & - \mu B_{T} & -\omega \cos 2\theta  & \omega \sin 2\theta \cr 
 \mu B_{T} & 0 & \omega \sin 2\theta & \omega \cos 2\theta
\end{pmatrix},
\label{hvac:2}
\end{equation}
where $B_T$ is the component of the magnetic field transverse to the neutrino momentum, and $\mu$ is the magnitude of the Majorana neutrino magnetic moment.  The Majorana transition magnetic moment matrix is antisymmetric and, in the case of two neutrino flavors, it is parameterized by only one number.

\section{Neutrino Magnetic Moments and Magnetic Fields inside Supernovae} 
\label{unknown:1}

When neutrinos are massive, weak interactions lead, at one-loop, to nonzero neutrino magnetic moments. Assuming only standard model interactions, if there were only one massive Dirac neutrino, its magnetic moment would be (for a recent overview, see \cite{Giunti:2008ve})
\begin{eqnarray}
\mu_{D} &=& \frac{3eG_{F} m_{\nu}}{8\pi^2\sqrt{2}}\left(1+{\cal O}\left(\frac{m_{\ell}^2}{M^2_W}\right)\right), \nonumber \\
&=& 3.2 \times 10^{-19} \left(\frac{m_{\nu}}{1~\rm eV}\right) \mu_{B},
\label{eq:mud}
\end{eqnarray}
where $m_{\ell}$ is a charged lepton mass, $M_W$ is the $W$-boson mass, and $\mu_B=5.788\times 10^{-9}$~eV/gauss is the Bohr magneton. For a neutrino mass of order a tenth of an eV, $\mu_D\sim 3 \times 10^{-20} \mu_{B}$. For more than one neutrino species, the transition magnetic moments are GIM suppressed with respect to the diagonal ones. Taking into account that all mixing angles in the lepton mixing matrix are large, the GIM suppression  factor is of order $m_{\tau}^2/M_W^2\sim 5\times 10^{-4}$.

For Majorana neutrinos only transition magnetic moments are allowed: the magnetic moment matrix is totally antisymmetric, $\mu^M_{ij}=-\mu^M_{ji}$ for $i,j=1,2,3$. The magnitudes of the Majorana neutrino transition electromagnetic moments depend on the neutrino Majorana phases (or, if $\mathcal{CP}$-invariance is conserved, on the relative $\mathcal{CP}$-parities of the different neutrino mass eigenstates). For generic Majorana phases one can estimate that standard model interactions lead to, assuming all neutrino masses have the same order of magnitude $m_{\nu}=0.1$~eV, $\mu^M_{ij}=-\mu^M_{ji}\sim 10^{-4}\mu_D\sim10^{-24}\mu_B$, where $\mu_D$ is given by Eq.~(\ref{eq:mud}).  For more details see, for example, \cite{Giunti:2008ve}. In theories beyond the standard model, much larger values for the Majorana neutrino transition magnetic moments are, of course, allowed. Our current understanding of neutrinos interactions constrains $\mu\lesssim 10^{-11}\mu_B$ \cite{Nakamura:2010zzi}, some nine orders of magnitude larger than $\mu_D$ above. 

The Hamiltonian Eq.~(\ref{hvac:2}) depends on the combination $\mu B_{T}$ and hence we need to discuss the expected magnetic fields inside the supernova explosion (we drop the transverse subscript $T$ on $B$ hereafter). We make use of a toy model for the magnetic field and assume it falls off like $1/r^{2}$, where $r$ is the distance from the center of the supernova, and assume a magnitude of $10^{12}$~gauss at 50~km, which is consistent with estimates in the literature (see, for example, \cite{Thompson:1993}) and might be considered, by some authors, to be conservative (see, for example, \cite{Maruyama:2012hf}):
\begin{equation}
B(r)=10^{12}\left(\frac{50~\rm km}{r}\right)^2~{\rm gauss}.
\end{equation}
It should be noted that the magnetic field in the supernova is probably toroidal in nature. Depending on the direction of neutrino propagation, it may decrease more gradually than assumed here, or even increase in magnitude at small radii~\cite{Obergaulinger:2005xd}. 

One needs also address the orientation of the magnetic field with respect to the supernova--earth direction, since it is not just the magnitude of the magnetic field but also its direction that affects the evolution of the neutrino fluxes. The magnetic field has no effect if it is parallel to the direction of neutrino propagation. Furthermore, we do not consider twisting magnetic fields. A dipolar magnetic field is sufficient to illustrate the importance of the transition magnetic moment in collective oscillations. Finally, it is also overly simplistic to consider a magnetic field which is fairly smooth without any turbulence, but the addition of turbulence can only make things more interesting. 

It will prove useful to define
\begin{equation}
(\mu_DB)_{\rm SM}(r) = \mu_{D}B_0\left(\frac{50~\rm km}{r}\right)^2\sim 1.9\times 10^{-10}\left(\frac{50~\rm km}{r}\right)^2~\left(\frac{\rm eV^2}{\rm MeV}\right),
\end{equation}
for $B_{0}=10^{12}$~gauss and $\mu_D = 3.2 \times 10^{-20} \mu_{B}$. We are interested in positions $200~{\rm km}\ge r\ge 50$~km. Our results will be presented in units of $(\mu_DB)_{\rm SM}$, including the $r^{-2}$ dependency. In these units, standard-model-like Majorana neutrino magnetic moments, assuming our estimates for the magnetic field are appropriate, correspond to $\mu B(r)=10^{-4}(\mu_DB)_{\rm SM}$. Magnetic field orientation issues may mean smaller effective $\mu B$ values, while magnetic fields larger than the ones considered here mean larger $\mu B$. If there is new physics beyond the standard model, much larger values --- many orders of magnitude larger! --- are allowed. 

\section{Results}
\label{sec:results}

When the neutrino magnetic moment effects are negligible, collective neutrino oscillations, for the values of the parameters discussed in Sec.~\ref{sec:formalism}, are depicted in Fig.~\ref{stable:1}. These results are well understood in the literature and can be qualitatively described, for the inverted mass hierarchy, as a ``spectral swap'' between $\nu_e$ and $\nu_x$ for energies between (roughly) 5~MeV and 18~MeV. 

\begin{figure}[!h]
\includegraphics[width=8.5cm]{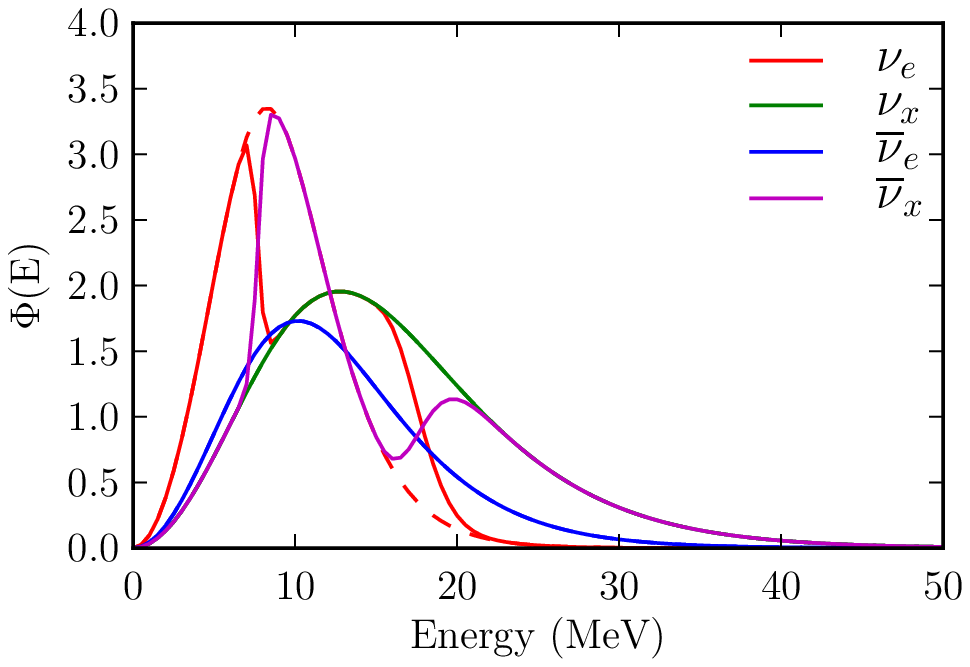}
\includegraphics[width=8.5cm]{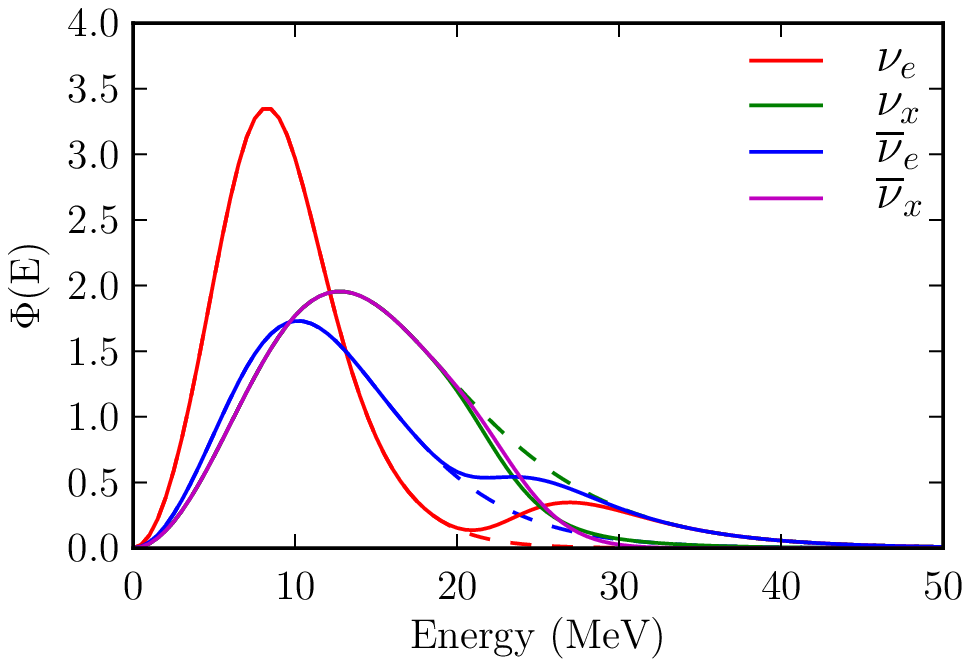}
\caption{Initial (dashed, $r=50$~km) and final (solid, $r=200$~km) $\nu_e, \bar{\nu}_e, \nu_x, \bar{\nu}_x$ neutrino fluxes as a function of the neutrino energy for $\theta=0$ and $\mu B (r) = 10^{-2}(\mu_{D} B)_{\rm SM} $ (see Sec.\ref{unknown:1}), for a normal (left) and inverted (right) neutrino mass hierarchy. See~\cite{youtube:list} for an animation of the radial evolution of the fluxes.}
\label{theta0:1}
\end{figure}

Technically, the addition of transition magnetic moment significantly increases the complexity of the differential equations, to the extent that it is very challenging to simply ``code'' them without the possibility of typographical errors. To circumvent this, we use SAGE \cite{sage} to automatically generate the C-code for the input functions. We performed multiple checks in order to verify that our results are, numerically, reliable, including exploring circumstances when either $\mu B$ or $\theta=0$ vanish. The former check has already been discussed. Eq.~(\ref{hvac:2}) reveals that a similar behavior is to be expected in the case $\mu B\neq 0$, $\theta=0$, except that, this time, the ``swapping states'' are $\nu_e$ and $\bar{\nu}_x$ (and their antiparticles).\footnote{In the upcoming figures, this is not immediately obvious due to the fact that $\nu_x$ and $\bar{\nu}_x$ have identical initial spectra.} Our results for $\theta=0$ and $\mu B (r) = 10^{-2}(\mu_{D} B)_{\rm SM}$ are depicted in Figure~\ref{theta0:1} and clearly show that the expectations are confirmed, this time for a normal neutrino mass hierarchy. Like in the standard case ($\mu B=0$, $\theta\neq 0$), one can verify the conservation of `lepton number' in the sense that $\Delta \nu_{e}$ is equal to $\Delta \bar{\nu}_{e}$. Indeed, this is one of the ``sanity checks'' we performed in order to establish the reliability of the numerical results presented here.  

We draw attention to the fact that $\mu B (r) = 10^{-2}(\mu_{D} B)_{\rm SM}$ is larger (by a couple of orders of magnitude) than the ``baseline'' standard model expectation for massive Majorana neutrinos discussed in Sec.~\ref{unknown:1}. Numerically, we find that $\mu B (r) = 10^{-4}(\mu_{D} B)_{\rm SM}$ does not lead, given the initial fluxes considered here, to a significant effect. We return to this issue in the next section. 

\begin{figure}
\includegraphics[width=8.5cm]{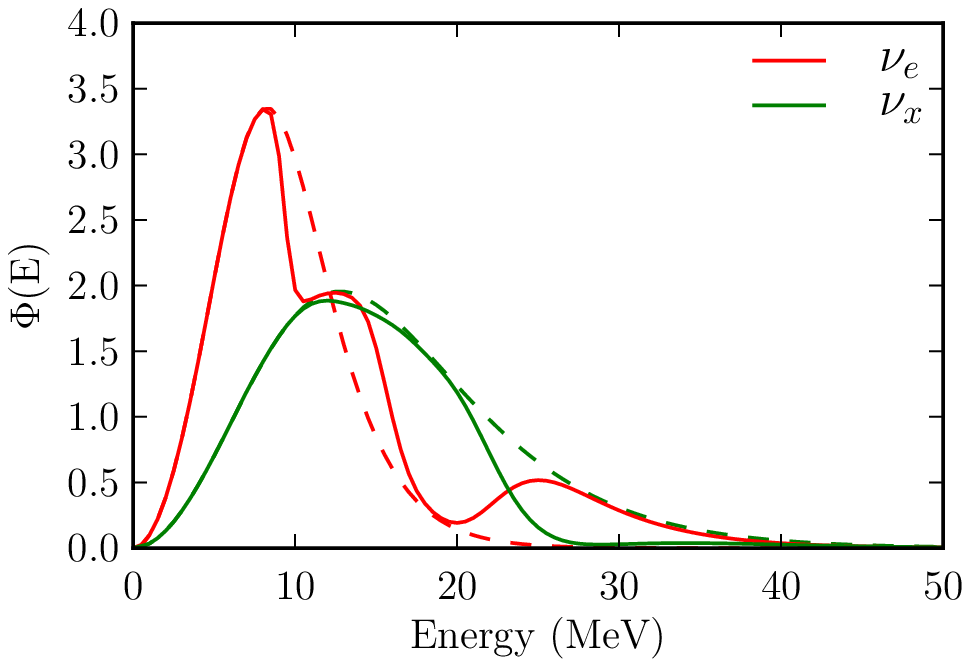}
\includegraphics[width=8.5cm]{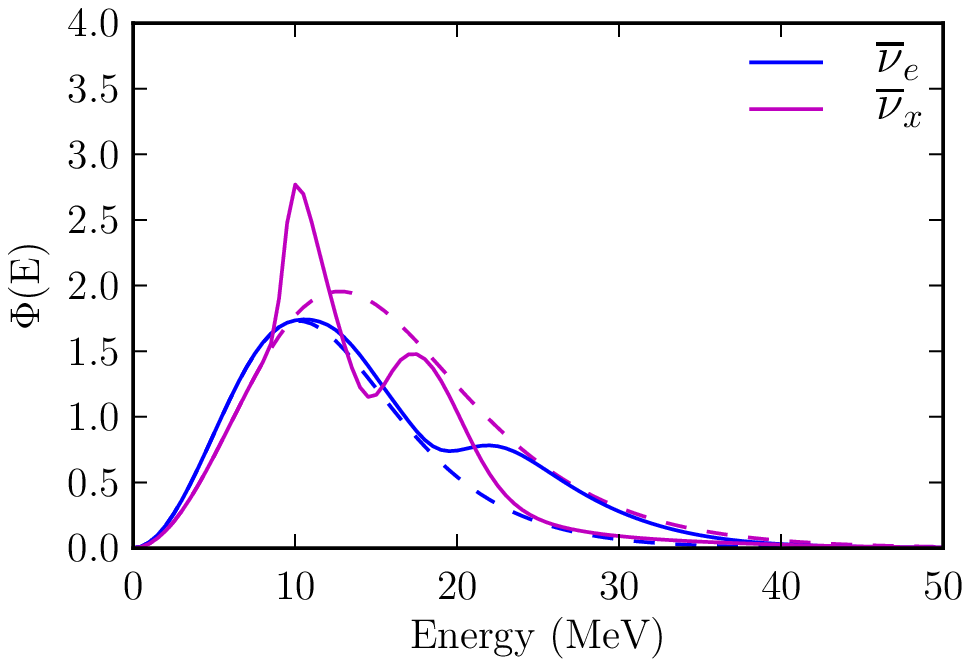}
\caption{Initial (dashed, $r=50$~km) and final (solid, $r=200$~km) $\nu_e, \bar{\nu}_e, \nu_x, \bar{\nu}_x$ neutrino fluxes as a function of the neutrino energy, including the effect of transition magnetic moment for neutrinos (left) and antineutrinos (right), for a normal mass hierarchy and $\mu B (r) = 10^{-2}(\mu_{D} B)_{\rm SM}$ (see Sec.\ref{unknown:1}). See~\cite{youtube:list} for an animation of the radial evolution of the fluxes.}
\label{nsimu:normal}
\end{figure}

\begin{figure}
\includegraphics[width=8.5cm]{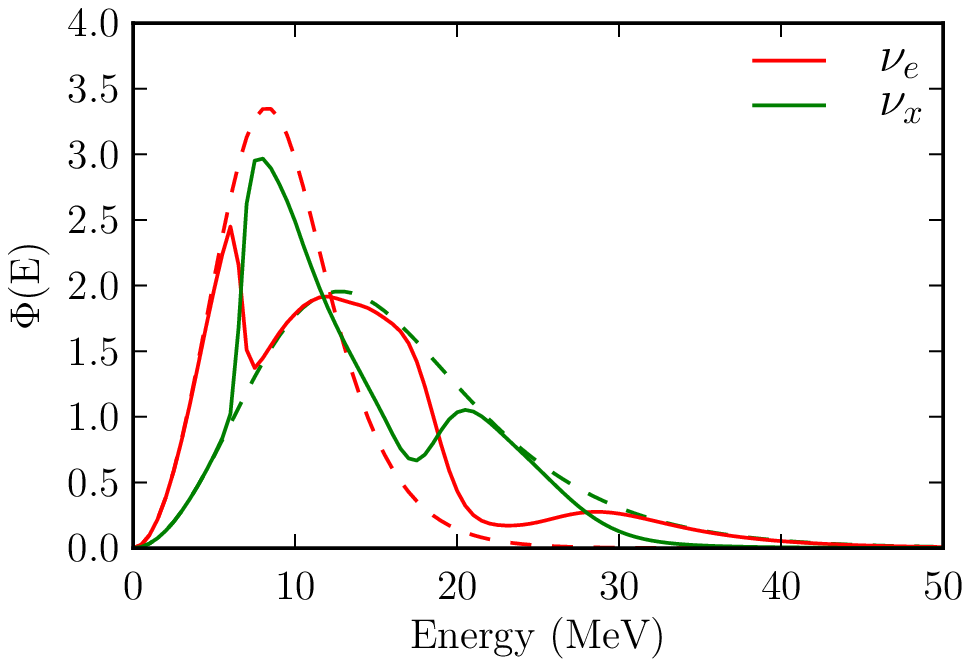}
\includegraphics[width=8.5cm]{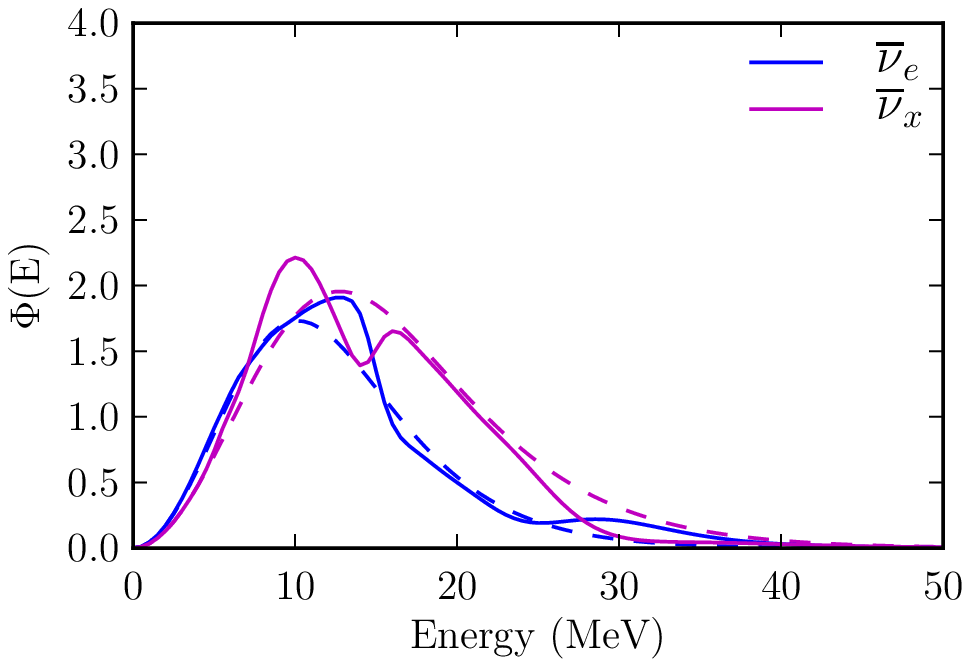}
\caption{Same as Fig.~\ref{nsimu:normal}, for an inverted neutrino mass hierarchy. See~\cite{youtube:list} for an animation of the radial evolution of the fluxes.}
\label{nsimu:inverted}
\end{figure}

Naively, one could anticipate that when both $\theta$ and $\mu B$ are nonzero, the combined collective oscillation effects would ``add up,'' and the final flux could be obtained by starting with the standard case ($\mu B=0$) and applying the logic mentioned earlier in order to estimate the effect of the nonzero $\mu B$. Our results for $\sin^2\theta=10^{-4}$, $\mu B (r) = 10^{-2}(\mu_{D} B)_{\rm SM}$ are depicted, for a normal and inverted neutrino mass hierarchy, in Figs.~\ref{nsimu:normal}, \ref{nsimu:inverted}. The figures reveal more structure in the final neutrino fluxes than naively anticipated. There are effects for both the normal and inverted mass hierarchies, and all flavor neutrinos and antineutrinos are affected. 
As we mentioned earlier, in the absence of the transition magnetic moment, the lepton number is conserved at each and every energy. Once the  transition magnetic moment is ``turned on,'' lepton number is violated at individual energies. Total lepton number (integrated over all neutrino energies), however, is conserved.

\begin{figure}
\includegraphics[width=8.5cm]{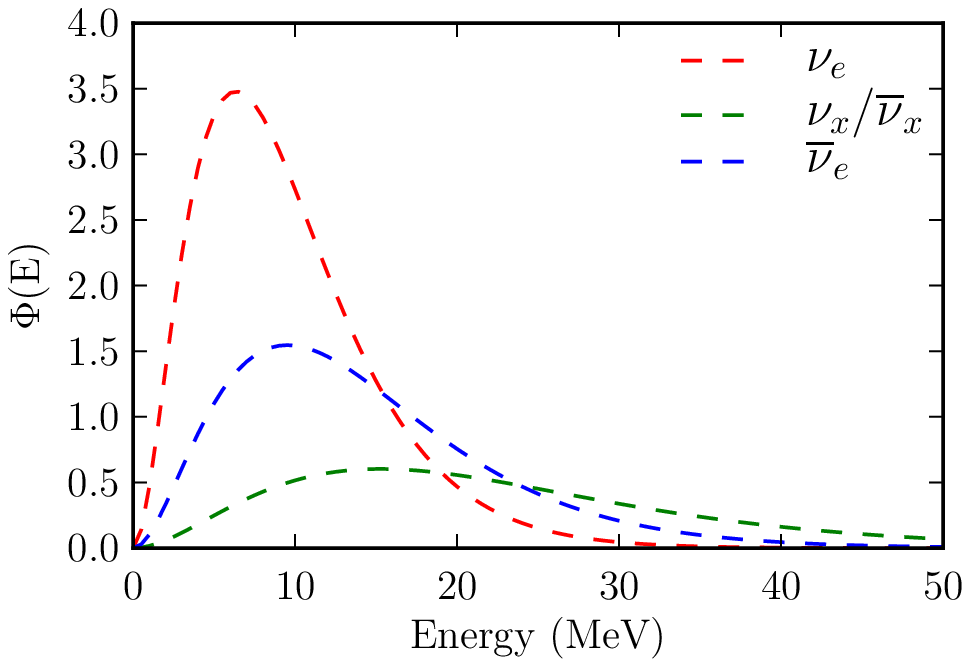}
\includegraphics[width=8.5cm]{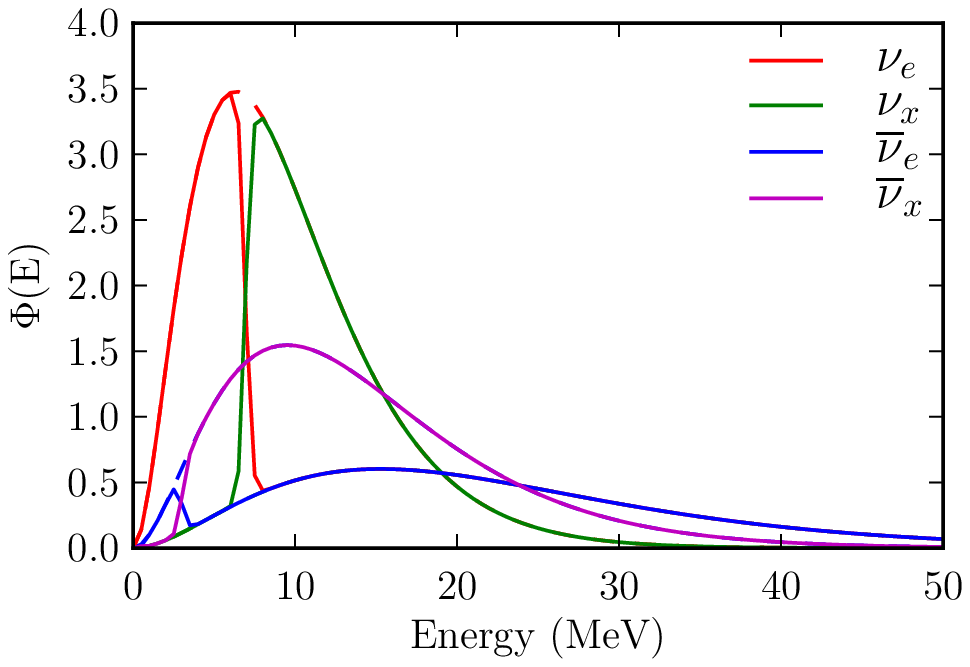}
\caption{Initial (left, $r=50$~km)  and final (right, $r=200$~km) $\nu_e, \bar{\nu}_e, \nu_x, \bar{\nu}_x$ neutrino fluxes as a function of the neutrino energy, assuming different initial neutrino spectra from the previous figures: higher $L_{\nu_{e}}:L_{\nu_{x}}$ ratio and $L_{\bar{\nu}_{e}}:L_{\nu_{x}}$ ratio, for $\mu B=0$ (no magnetic moment effect) and an inverted neutrino mass hierarchy. For these initial fluxes, collective oscillations are virtually absent for the normal neutrino mass hierarchy. See~\cite{youtube:list} for an animation of the radial evolution of the fluxes.}
\label{lisi:stable}
\end{figure}

\begin{figure}
\includegraphics[width=8.5cm]{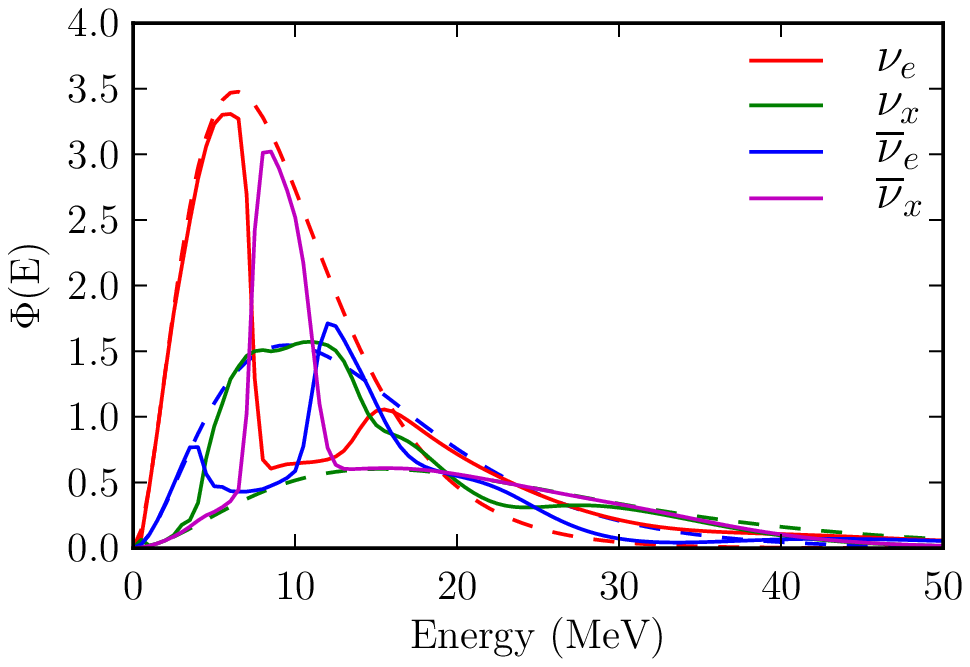}
\includegraphics[width=8.5cm]{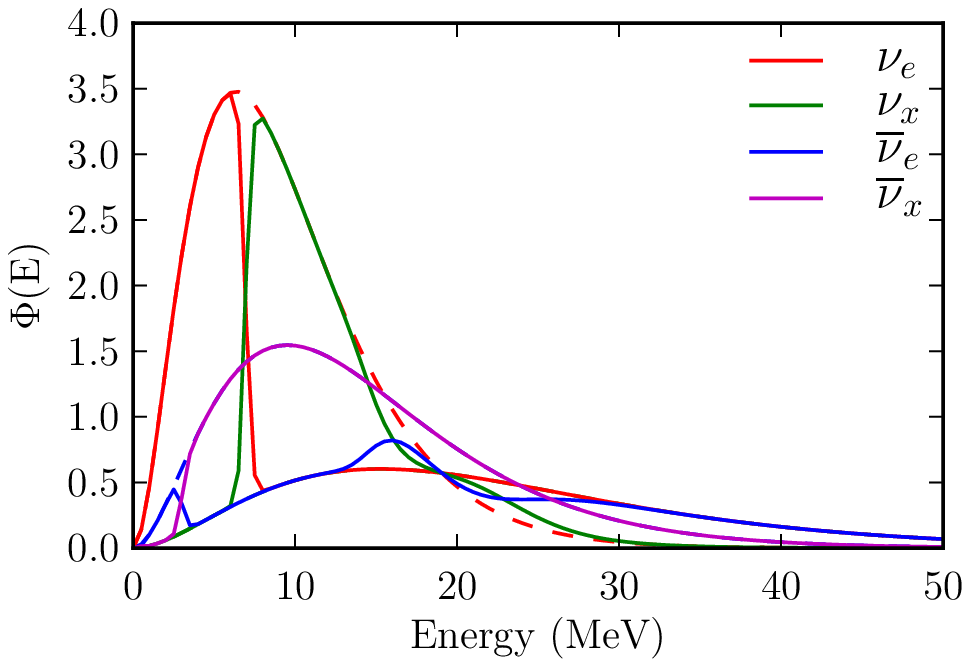}
\caption{Same as Fig.~\ref{lisi:stable}, for $\mu B(r)=10^{-2} (\mu_{D} B)_{\rm SM}$ and a normal (left) or inverted (right) neutrino mass hierarchy. See~\cite{youtube:list} for an animation of the radial evolution of the fluxes.}
\label{lisi:flux}
\end{figure}

Different initial fluxes lead to quantitatively different results. We consider initial fluxes, depicted in Figure~\ref{lisi:stable}(left) (see \cite{Fogli:2007bk}) where the ratios $L_{\nu_{e}}:L_{\nu_{x}}$ and $L_{\bar{\nu}_{e}}:L_{\nu_{x}}$ are significantly higher than the ones in computed in \cite{Keil:2002in}, and repeat the computations performed above. These results, depicted in  Figure~\ref{lisi:stable}(right)\footnote{For this flux and a normal hierarchy, spectra are virtually unaffected by collective oscillations.} and Figure~\ref{lisi:flux}, are quantitatively different from the ones in Figs.~\ref{stable:1}, \ref{nsimu:normal}, and \ref{nsimu:inverted}. Here, we see that the nonzero magnetic moment effects are much more pronounced --- there are multiple splits --- in the case of a normal mass hierarchy, while in the case of an inverted one the effect of the nonzero magnetic moment is virtually absent. 

The comparison of Figs.~\ref{lisi:stable}(right), \ref{lisi:flux} with Figs.~\ref{stable:1}, \ref{nsimu:normal}, and \ref{nsimu:inverted} illustrates the challenge of disentangling ``particle physics'' effects from ``astrophysics effects.'' For the two sets of figures the neutrino parameters are, pairwise, the same, while the initial neutrino fluxes are different. The structure in the final fluxes is clearly different. 
Regardless of the initial neutrino spectra, however, our results allow one to conclude that modest values for the neutrino magnetic moment, in the region of the supernova explosion where collective effects are dominant ($r\lesssim 200$~km), can significantly modify the energy spectrum of neutrinos of different flavors, and that, once one combines the ``standard'' $\theta_{13}$ effect with a nonzero magnetic moment, spectral splits --- sometimes multiple splits --- are expected for both neutrino mass hierarchies.

In order to relate theoretical fluxes with potential observables at terrestrial experiments, it is necessary to further ``evolve'' the neutrinos from $r=200$~km to $r\to\infty$. Unfortunately, the standard matter effects, which dominate outside the ``collective oscillations'' region, cannot be treated in a satisfactory manner in the two-flavor approximation, as the multiple level crossings play an important role at larger radii. This by itself is a very interesting topic~\cite{Akhmedov:1996ec,EstebanPretel:2007ec}, and has to be combined with the study of collective oscillations in order to draw any conclusions regarding the observable signatures due to the transition magnetic moment. The three flavor analysis of this problem, which is beyond the intentions of this manuscript, would take us a step closer towards possible observable signatures in the final flux. It contains additional intricacies, including potentially three distinct transition magnetic moments. A detailed analysis of the three flavor effect in collective oscillation, and its interplay with the standard MSW effect in supernovae at larger radii, is work in progress which we hope to report in the near future.

\section{Conclusion}
\label{sec:conclusion}

We have studied the effect of nonzero Majorana transition magnetic moments on neutrino collective oscillations in the interior of supernova explosions. Our main results are illustrated in Figures~\ref{theta0:1}, \ref{nsimu:normal}, and \ref{nsimu:inverted}, and in \cite{youtube:list}. We find that, even for modest values of the neutrino magnetic moment and the magnetic field inside the supernova, nontrivial effects are expected. These are similar in character and magnitude to effects of nonzero leptonic mixing angles, which have been thoroughly studied in the literature. The interplay between ``magnetic moment effects'' and ``flavor mixing effects'' is also interesting and may lead to the presence of nontrivial structure in the neutrino flavor spectra for both the normal and the inverted neutrino mass hierarchies. 

We find that nontrivial magnetic moment effects occur for $\mu$ values that are only a couple of orders of magnitude larger than the standard model expectations, assuming no new physics beyond light neutrino Majorana masses. It is, arguably, virtually impossible to imagine another potential observable --- astrophysical or terrestrial --- that is, in practice, sensitive to such tiny neutrino magnetic moments. Our result is dependent on the very simple, albeit mostly conservative, assumptions we made regarding the magnitude, form, and radial dependency of the magnetic field in the supernova core. The fact that one is sensitive to unprecedentedly small values of the neutrino magnetic moment is, however, robust under changes of the hypothesis regarding magnetic fields.

We have restricted our discussion to the propagation of neutrinos inside the region where collective oscillations dominate ($r\lesssim 200$~km). Supernova neutrino spectra, when observed in terrestrial experiments, will be further ``processed'' by neutrino propagation further downstream within the supernova, including regions where the ``standard'' matter effects are strong, and potential propagation through the Earth. Throughout, if the neutrino magnetic moments are much larger than standard model expectations (and much larger the the values considered here) other magnetic moment effects are expected outside the region under consideration here. Disentangling all of these effects from the magnetic-moment induced collective effects discussed here is a formidable task we have not addressed here at all. We do, however, point out that the magnetic moment effects considered here may partially mimic effects characteristic of the wrong neutrino mass hierarchy, and that the entire picture is also sensitive to the initial neutrino flavor spectra inside the supernova core. 

Disentangling the physics contained in measured supernova neutrino spectra (which will happen in the future, hopefully at multiple detectors) is a well known challenge. We point out that neutrino electromagnetic properties are, potentially, an important component of this ``entanglement.'' The physics opportunities, on the other hand, are very exciting. For example, if astrophysical uncertainties are under control, and if the neutrino mass hierarchy is known, it is possible that, by observing supernova neutrinos, we may be able to determine that the neutrino magnetic moments are nonzero. Furthermore, if Majorana magnetic moments are qualitatively different from Dirac ones ({\it cf.} discussion in the last paragraph), it is possible we may be able to indirectly infer that neutrinos are Majorana fermions just by observing effects of neutrino magnetic moments from supernovae. Such an inference would be completely independent from other observables sensitive to the nature of the neutrino.

Our main goal was to point out that even very small neutrino Majorana transition magnetic moments can significantly impact the flavor structure of supernova neutrinos. Along the way, we have made several simplifying assumptions that need to be revisited if one is to obtain more robust quantitative results. In more detail, we assume the existence of only two neutrino flavors and have restricted our analysis to the single angle approximation. There may be other interesting features of collective oscillations we have missed due to these approximations. In the absence of neutrino magnetic moments, for example, the inclusion of three-flavor effects (see \cite{Dasgupta:2008cd,EstebanPretel:2007yq,Dasgupta:2007ws,Duan:2008za,Fogli:2008fj,Friedland:2010sc,Dasgupta:2010cd,Duan:2010bf}) and multi-angle analyses (see \cite{Sawyer:2008zs,Duan:2008eb,Duan:2008fd,Duan:2010bg}) have revealed some interesting effects not properly captured by the two-flavor, single-angle approximation. 

Throughout, we have restricted our analysis to Majorana neutrinos. If the neutrinos are Dirac fermions the problem is both technically more involved and qualitatively different. If the neutrinos are Dirac fermions, the magnetic moment interacting with the local magnetic field leads to ``active--sterile'' oscillations (where the new right-handed neutrino degrees of freedom act, for all practical purposes, as sterile neutrinos) and the ``two-flavor'' approximation is, in reality, a ``four-flavor'' problem. Furthermore, Dirac neutrinos can also carry a diagonal magnetic moment and the number of free parameters is, in the two-flavor--case, three (as opposed to one in the Majorana case). Qualitatively, using previous results as guidance, we don't expect any sizable effects. The reason is simple. Collective oscillations obey certain conservation laws and the fact that the initial flux of sterile neutrinos is zero implies that no ``swaps'' can take place --- there are no sterile neutrinos around to trade places with the active ones! Preliminary results indicate that this naive picture is indeed realized. 

\appendix
\section{The Self-interaction Hamiltonian}
\label{app:1}
At the quantum field level, the neutrino--neutrino interaction Hamiltonian density is
\begin{equation}
{\cal H}_{self} = \sqrt{2}G_{F} \sum_{f,f'}(\bar{\psi}^{f}\gamma^{\mu}\psi^{f} )(\bar{\psi}^{f^{\prime}}\gamma_{\mu}\psi^{f^{\prime}}),
\end{equation}
where $f,f'=e,x$ and 
\begin{equation}
\psi^{f}(x) = \int \frac{d^{3}\mathbf{p}}{(2\pi)^{3}}\frac{1}{\sqrt{2 E_{p}}} \left(a^{f}_{p}u(\mathbf{p})e^{-i px} + b^{f\dagger}_{p}v(\mathbf{p})e^{i px} \right).
\end{equation}
The creation and annihilation operators were defined in Eq.~(\ref{creation_op}). The related field $\psi^{\bar{f}}(x)$ can be defined as $\mathcal{C}\psi^{f}(x)\mathcal{C}$. While we are dealing with Majorana particles, we can still safely consider fields made up of different creation and  annihilation operators corresponding, respectively, to left-hand and right-handed particles.

For sake of presentation, we derive a particular element of $H_{self}$,\footnote{All other ones can be derived in exactly the same way.} namely $(H_{self})_{e\bar{x}}$ which, (using $d^{3}q \equiv d^{3}\mathbf{q}/(2\pi)^3 2E_{q}$) can be written as,
\begin{eqnarray}
(H_{self})_{e\bar{x}}(p) &=& \sqrt{2}G_{F} \int d^{3}p^{\prime}d^{3}qd^{3}x
\left< \nu_{\bar{x}}(\mathbf{p}^{\prime}) \left| \bar{\psi}^{\bar{x}}\gamma^{\mu} \psi^{\bar{x}} \right|\mathbf{q}\right>\nonumber\\ 
& \times & \left< \mathbf{q} \left| \bar{\psi}^{e}\gamma_{\mu} \psi^{e} \right|\nu_{e}(\mathbf{p})\right>,\nonumber\\
&=& - \sqrt{2}G_{F} \int d^{3}p^{\prime}d^{3}qd^{3}x
\left< \nu_{\bar{x}}(\mathbf{p}^{\prime}) \left| \bar{\psi}^{\bar{x}}\gamma^{\mu} \psi^{e} \right|\nu_{e}(\mathbf{p})\right> \nonumber\\
& \times &\left< \mathbf{q} \left| \bar{\psi}^{e}\gamma_{\mu} \psi^{\bar{x}} \right|\mathbf{q}\right>.
\label{V:mat}
\end{eqnarray}
Here, the one-particle neutrino states of well-defined momentum are defined as
\begin{equation}
|\nu_f(\mathbf{p})\rangle = \sqrt{2E_{p}}a^{f\dagger}_{p}|0\rangle,
\label{cre:op}
\end{equation}
for $f=e,x$. 
These eigenstates are normalised according to
\begin{equation}
\langle \mathbf{p}^{\prime} f^{\prime}| \mathbf{p},f\rangle = 2E_{p}(2\pi)^{3}\delta^{(3)}(\mathbf{p}-\mathbf{p}^{\prime})\delta^{ff^{\prime}}.
\end{equation}
The same expressions with $a\leftrightarrow b$ apply for antineutrino states. $|\mathbf{q}\rangle$ are a complete set of states (including neutrinos and antineutrinos, and all flavors). In Eq.~(\ref{V:mat}) we have used the Fierz identity in the second line, so that we can see that for the second matrix element the only creation and annihilation operators that survive are $b_{e}a^{\dagger}_{x}$ and $a^{\dagger}_{e}b_{x}$ with a relative minus sign between them. Using the elements of the density matrix,
\begin{eqnarray}
\left< \mathbf{q} \left| \bar{\psi}^{e}\gamma^{\mu} \psi^{\bar{x}} \right|\mathbf{q}\right> \propto \rho_{{e}\bar{x}} \langle \nu_{e} |J^{\mu}_{\bar{x}e} a_{e}^{\dagger} b_{x} | \nu_{\bar{x}} \rangle + \rho_{x\bar{e}} \langle \nu_{x} | J^{\mu}_{\bar{e}x} b_{e} a_{x}^{\dagger} | \nu_{\bar{e}} \rangle,
\end{eqnarray}
A relative negative sign between the two terms arises from $J^{\mu}_{\bar{x}e}=\bar{u}\gamma^{\mu}v=-J^{\mu}_{\bar{e}x}$, as these are $\mathcal{CP}$ conjugates of each other. Finally, we use the standard technique of summing over the spins by taking the trace which leads to a dependency of $(H_{self})$ on the angle between the momenta of the neutrinos,
\begin{eqnarray}
(H_{self})_{e\bar{x}} &=& \sqrt{2} G_{F} \int d^{3}\mathbf{q} \left(\rho_{x\bar{e}} - \rho_{e\bar{x}} \right)\left(\frac{p^{\mu}}{E_{p}}\frac{q_{\mu}}{E_{q}}\right), \\
                &=& \sqrt{2} G_{F} \int d^{3}\mathbf{q} \left(\rho_{x\bar{e}} - \rho_{e\bar{x}} \right)\left(1-\cos(\theta_{pq})\right).
\end{eqnarray}
The potential derived above can be averaged over the angle between relative velocities $\theta_{pq}$ to obtain an average potential per particle in the thermal bath. The potential experienced by a particle traveling through the thermal bath is proportional to the number of particles in the bath.
The above equation can be written in the form presented as Eq. \ref{h:self} with the help of $\rho^{c}$: $\rho_{x\bar{e}}=\rho_{e\bar{x}}^{c*}$.

The matrix elements of $(H_{self})_{e\bar{x}}$ can be obtained from the expression for $(H_{self})_{ex}$ in \cite{Sigl:1992fn} as follows. Every time we replace a neutrino by its antiparticle, the corresponding current (see Eq.~(\ref{V:mat})) acquires a relative negative sign. This means that, in order to derive the full self-interaction Hamiltonian, the number of matrix elements that need to be calculated is very small. This sign flip is also the reason why the coupling constant matrix has the form $G=\mathrm{diag}(1,1,-1,-1)$, in Eq.~(\ref{h:self}). This form of the coupling constant matrix ensures that there is a negative sign whenever any two flavors in the equations of motion are replaced by their charge conjugate. 

\acknowledgments{
The authors are indebted to Alex Friedland for many useful discussions and comments on the manuscript, and to Basu Dasgupta for careful comments on the manuscript and plenty of encouraging words. We also thank Georg Raffelt for questions and comments. This work is sponsored in part by the DOE grant \#DE-FG02-91ER40684.
}
\bibliography{magmom2.bib}
\bibliographystyle{h-physrev}
\end{document}